# Optical spectroscopy of excitons in ReS$_2$ monolayers grown by chemical vapor deposition

Solomon Ojo[1], Juwon Onasanya[1], Morad Benamara[2], Bothina Hamad[3], and M. O. Manasreh[4, a)]

[1]Material Science and Engineering program, University of Arkansas, Fayetteville, AR 72701

[2]Institute of nanotechnology, University of Arkansas, Fayetteville, AR 72701

[3]Physics Department, University of Arkansas, Fayetteville, AR 72701

[4]Department of Electrical Engineering and Computer Science, University of Arkansas, Fayetteville, AR 72701

## Abstract

Monolayers of ReS$_2$ were grown by a chemical vapor deposition technique on SiO$_2$/Si substrates and investigated at room temperature by using μ-Raman, μ-photoluminescence (PL) and absorbance spectroscopies. The Raman scattering spectrum exhibits several phonon modes that were confirmed by the computation analysis based on the density functional theory. The ReS$_2$ structural integrity was confirmed by using the XRD and the energy dispersion spectroscopy that was obtained by the scanning electron microscopy. Photoluminescence spectra show excitons related to interband transition in thin monolayer flakes and bulk-like structures. Additionally, a sharp PL line located at 1.2620 eV with a second harmonic peak at 2.5420 eV were observed and explained in terms of interband excitonic transitions originated within the rhenium 5d orbit. An optical absorbance spectrum with an exciton peak around 1.4660 eV was obtained for an assembly of ReS$_2$ flakes grown on silica substrate.

Keywords: ReS$_2$, Interlayer excitons, second harmonic generation, phonons, PL, absorption.

[a)]The corresponding author: manasreh@uark.edu

**Introduction**:

Two-dimensional (2D) transition metal dichalcogenides (TMDs) attracted significant attention due to their remarkable electronic, optical, and mechanical properties.[1] Another noticeable property of this class of materials is that they can be formed in heterojunctions where the monolayers are held together by van der Waals forces.[2] Rhenium disulfide (ReS$_2$) is one of these two-dimensional (2D) materials that is characterized by a distorted octahedral (1T´) structure with a space group $P\bar{1}$ and C$_i$ point group, which has only the inversion symmetry. The reduced symmetry of this material leads to an anisotropic electronic behavior and interesting excitonic transitions.[3] One particularly intriguing phenomenon in TMDs is the presence of second-harmonic generation (SHG) that is a nonlinear optical process in which light interacts with a polar material and generates light at twice the incoming frequency. The second harmonic generation is forbidden in centrosymmetric materials. However, a broken symmetry at the atomic level can enable it in certain TMDs, including ReS$_2$,[4] which is a process observed in many 2D materials, such as WSe2.[5] Furthermore, interlayer excitonic transitions are reported in many 2D heterojunction materials, such as MoSe$_2$-WSe$_2$,[6,7] homo-bilayers,[8,9] monolayers ReS$_2$,[10] and even bulk ReS$_2$.[11]

Interband excitons in ReS$_2$ can also be originated from nonbonding Re 5d $t_{2g}$ to 5d $t_{2g}^*$ transitions.[12] These orbital excitons are observed as sharp lines in reflectance measurements.[12] Excitons in low dimensional materials exhibits two advantages over their bulk counterparts. The binding energy of excitons in 2D materials is usually higher than the binding energy of excitons in bulk materials. The other advantage is the lifetime of exciton in 2D materials is longer than that in the bulk materials. These two advantages put the 2D materials in the forefront of device applications that may operate at room temperature. By exploring the unique band structure and atomic arrangement of ReS$_2$, one may expect to observe applications and phenomena useful for optoelectronic applications, such as photodetectors spanning the spectral range of infrared to visible[13,14,15] with significant high values of figures of merit.

In this article, we report on the optical properties of excitons in $ReS_2$ material that was grown by a chemical vapor deposition (CVD) technique. Flakes were grown on either $SiO_2$/Si or silica substrates. The integrity of the crystal structures was confirmed with both the scanning electron microscopy (SEM) and the energy dispersion spectroscopy (EDS). The phonon modes were measured by using μ-Raman scattering method and theoretically confirmed by the density-functional perturbation theory[16] (DFPT). The photoluminescence (PL) spectra were measured for samples grown at different temperatures. The optical absorbance spectra were recorded for samples grown on silica substrates.

**Experiment and theory**:

The CVD system consists of a Lindberg/Blue M furnace with quartz tubing and several mass flow controllers. Pure solid sulfur (S) was used. Ammonium perrhenate ($NH_4ReO_4$) powder is the precursor for rhenium (Re). The Re precursor was obtained from Sigma-Aldrich with a purity of 99.9%. The growth temperature was set between 500 – 700 °C. Both μ-Raman and μ-PL spectra were recorded using Horiba LabRam HR spectrometer with a HeNe laser (632.81 nm) was set for the phonon measurements and a Cobolt laser (472.92 nm) was used for the PL measurements. The optical absorbance spectra were collected using Cary 500 spectrometer. All measurements were performed at room temperature. The computational calculations of the phonon modes were obtained by employing the DFPT utilizing the ESPRESSO software package.[17]

**Results and discussions:**

Several samples were grown at different temperatures and investigated. An example of samples grown at 700°C on $SiO_2$/Si substrate is shown in Fig. 1(a) where the typical equilateral triangular shape of the flakes is dominant. The triangular side dimension of these flakes is shown in Fig. 1 (b) as 8.62 µm. The equilateral triangular shape of many 2D TMD was also reported in other studies.[18, 19] It is noted that these flakes are easy to peel off from the surface of the substrate indicating that their thickness is most likely a few monolayers as confirmed by the PL measurements. The atomic percentage ratio of Re:S was obtained from the EDS measurement as 1:2, which confirms the chemical structure of $ReS_2$. The density of the flakes was found to be higher when are grown on silica substrates with smaller flake sizes as shown in Fig. 1 (c). The XRD pattern of $ReS_2$ is shown in Fig. 1(d), which is in good agreement with results obtained by others.[20, 21]

The Raman shift spectrum of $ReS_2$ is obtained for samples as described in Fig. 1 and the results are shown in Fig. 2. The spectrum in this figure exhibits 17 peaks related to $A_g$-symmetry phonon modes. The "A" representation in $A_g$ symmetry refers to the rotation symmetry around the principal axis, while "g" corresponds to the inversion symmetry. The result of the computational analysis using the DFT is listed in Table I along with the peak positions of the experimental Raman modes. The discrepancy between the experimental measurements and the theoretical prediction is less than 4.0%. These phonon modes resulted from a combination of bulk and monolayers phonons.[19,22,23] The number of phonon modes (17 modes) observed here is in good agreement with the number of reported modes (18 modes).[22] The current results indicate that peak number 9 in Fig. 2 is degenerate and composed of two close modes where the monolayer mode is more dominant over the bulk mode.

The optical properties of ReS$_2$ are highly anisotropic in the basal plane,[24, 25] including the PL spectroscopy.[26] An excitonic transitions are observed in this study as shown in Fig. 3. The µ-PL spectrum shown in Fig. 3 (a) is obtained for a flake identical to that in Fig. 1(b). This peak, located at E = 1.887 eV (0.657 µm), is an excitonic peak generated from interband transition (electron-hole pair) within the same monolayer(s). The energy position and the line shape of this excitonic peak indicates that the flake is a few monolayer thick and that these monolayers exhibit direct band gap energy. This exciton peak energy is higher than the reported PL peak of ~1.6 eV, but compatible with the reflectance measurement for one monolayer.[25] It is also in good agreement with the PL results reported for ReS$_2$/2D perovskite heterojunction.[27] The broad PL peak in Fig. 3 (a), labeled E$_{bulk}$, is identical to the PL peak observed for bulk materials that obtained from a commercial source. Furthermore, a sharp peak is observed at 1.2616 eV (0.9830 µm). This PL peak is expanded and shown in the inset of the figure. It is composed of two peaks labeled IX$_1$ and IX$_2$, which are assigned to orbital transitions originated from nonbonding Re 5d $t_{2g}$ to 5d $t_{2g}^*$ transitions.[12] These two peaks are separated by about 3.3 meV. Closely related two sharp exciton peaks were observed by the electrolyte electroreflectance measurements[12] in ReS$_2$. However, the energy positions of these two excitons are smaller than the reported energies for the two nonbonding excitons n ReS$_2$ by about 0.20 eV.[12]

The PL spectra were collected for several flakes from the sample shown in Fig. 1(a). These spectra are plotted in Fig. 3 (b). While it is difficult to measure the thicknesses of the flakes, it clear from the PL spectrum that there is a myriad variation in the thicknesses of the flakes. It is ranging from a few layers with a triangular shape (see the inset of the spectrum labeled 1) to a bulk-like flakes as shown in the inset of the spectrum labeled bulk. The energies of the PL peaks are shown in the figure, which is obvious that the energy of the peaks is decreased as the thickness of the flakes is increased. The exciton peak observed as sharp peak in the spectrum labeled bulk in Fig. 3 (b) is also observed in small flakes, less than 2 µm cross, and plotted at the bottom of the figure that is labeled SHG.

The bottom spectrum in Fig. 3 (b) is expanded and plotted in Fig. 4 (a) for a flake from a sample grown at 500 °C on SiO$_2$/Si substrate. An image of the sample is shown as the inset in the figure. Two sharp peaks were observed at $E_1 = 1.2616$ and $E_2 = 2.5418$ eV. Both peaks are enlarged and shown in the insets. They both exhibit the same line shape with two peaks that are labeled IX$_1$ and IX$_2$ in the inset of Fig. 3 (a). Recent computations based on the DFT and pseudopotential theory for Re 5d $t_{2g}$ to 5d $t_{2g}^*$ transition in a slab (one monolayer) of ReS$_2$ is reported[28] with a gap of about 1.40 eV. This gap is comparable with the computed present bang gap of 1.43 eV as shown in Fig. 4 (b) where the electronic band structure and the projected density of states for the Re d-orbital along with the S p-orbital are plotted. The presence of partially filled d-orbitals in rhenium and their role in the valance band of ReS$_2$ make the transitions between these d-orbitals highly probable.[29] The present result of $E_1 = 1.2616$ eV follows this theoretical trend since it is larger than the band gap of bulk ReS$_2$, but thicker than a monolayer. It is very interesting to notice that the peak $E_1$ in Fig. 4 (a) is similar to the results reported for interlayer exciton in MoS$_2$[30] where two excitonic peaks were resolved under an applied electric field. Interlayer exciton formation was the subject of many investigations in 2D TMD materials.[7, 30 -33] However, the present results in Fig. 4 (a) (peak labeled $E_1$) is in line with nonbonding exciton transitions[12] from Re 5d $t_{2g}$ to 5d $t_{2g}^*$ atom rather than the interlayer excitonic peak. The peak labeled $E_2$ in Fig. 4 (a) occurs at an energy twice as that of $E_1$. The full width at have maximum (FWHM) of $E_2$ is about twice as that of the $E_1$ peak. This $E_2$ transition is assigned here as the second harmonic generation of $E_1$. The separation between these two peaks is $\Delta E = E_2 - 2E_1 = 2.5418 - 2 (1.2616) = 18.6$ meV. This is translated to $\Delta k=150.02$ cm$^{-1}$, where $\Delta k$ is the phase matching term[34] for which the coherent length[35] is $L_c= 2\pi/\Delta k = 0.42$ mm.

An excitonic transition is observed in the optical absorbance spectrum measured at room temperature as shown in Fig. 5 for ReS$_2$ sample grown on silica at 500 °C. An image of this sample is shown in Fig. 1 (c). Silica substrate is transparent in the near infrared – visible region, which allows the observation of optical transmission/absorption in case of ReS$_2$. A clear exciton peak is observed in the absorbance spectrum around $E_{abs}$ = 1.466 eV. This excitonic peak energy is higher than the energy (1.260 eV) of exciton PL energy observed in bulk-like ReS$_2$ as shown in Fig. 3 (b). As a comparison, a PL spectrum is plotted in Fig. 5 with a peak at $E_{PL}$ = 1.887 eV to indicate that thinner flakes possess larger band gaps. The exciton observed in the absorbance spectrum implies that the flakes, which are grown on silica substrate at lower temperature (500 °C), are composed of several monolayers, but not thick enough to be considered bulk-like flakes. Similar absorption measurement performed at 80 K were reported for bulk ReS$_2$ with an exciton peak located around 1.55 eV, but an excitonic peak was not present in a spectrum measured at room temperature.[36] The presence of excitonic transitions in the PL and absorbance spectra at room temperature is another proof for the exciton large binding energy in this class of materials, which make them resilient for device applications.

**Conclusion**:

Optical spectroscopies were used to investigate exciton transitions in ReS$_2$ flakes grown by a chemical vapor deposition technique. The crystal structure of the monolayers flakes is confirmed by using the XRD, EDS, and Raman scattering measurements. Phonon modes were calculated using the density functional theory. The excitonic transitions probed by the μ-PL were found to be sample dependent, which are shifted toward higher energy as the thickness of flakes is decreased. Additionally, optical absorbance spectra show the presence of excitons in flakes grown on silica substrates. An excitonic transition with a sharp energy peak in the PL spectra is observed and interpreted as being a transition within the nonbonding Re 5d orbital. A second harmonic generation peak with an energy twice the energy of the nonbonding exciton in Re 5d orbit is observed in thin monolayers flakes.

**Acknowledgement**: The authors would like to thank Arkansas High Performance Computing Center (AHPCC), University of Arkansas, for the computing resources.

**Funding**: The author(s) declare financial support was received for the research, authorship, and/or publication of this article. BH is supported by the MonArk Quantum Foundry that is funded by the National Science Foundation Q-AMASE-i program under NSF Award No. DMR-1906383.

**Authors contribution**: Solomon Ojo ran the CVD growth of the samples, made part of the optical measurements, XRD, and part of the theoretical computations. Juwon Onasanya helped with CVD growth and run part of the experimental measurements. Morad Benamara made the SEM and ESD measurements. Bothina Hamad supervised the theoretical computation, wrote part of the paper, and edit the paper. M. O. Manasreh supervised the experimental measurements and wrote most of the text including the final editing.

**Corresponding author**: M. O. Manasreh, Department of Electrical Engineering and Computer Science, University of Arkansas, Fayetteville, AR  72701.
http://Orcid.org/000-0002-2252-6543
Email:  manasreh@uark.edu

conference, Sinaia, Romania, 11 – 14 October 1995.

HTTP://DOI.ORG/10.1109/SMICND.1995.494915

**Figure captions**

Fig. 1. (a) An image of scanning elecron microscopy (SEM) for ReS$_2$ grown on SiO$_2$/Si substrates at 700 °C. Most of the flakes are triangular in shape. (b) The equilateral triangular side-dimention is 8.62 µm for most flakes as shown in the SEM image. (c) An image of ReS$_2$ flakes grown on silica substrate at 700 °C. (d) A typical XRD spectrum observed for ReS$_2$ samples.

Fig. 2. Micro-Raman spectrum measured at room temperature for an equilateral flake of ReS$_2$ grown on SiO$_2$/Si substrate. The peaks are labeled and tabulated in Table I alongside the calculated phone modes. Peak number nine is asymmetrical and it is composed of bulk and monolayer phonon modes

Fig. 3. (a) Photoluminescence spectrum of ReS$_2$ flake showing a single peak at 1.887 eV and a broad bulk-like peak, labeled E$_{bulk}$. Superimposed on the broad peak is another sharp peak at 1.2616 eV. This peak is assigned to a nonbonding Re 5d $t_{2g}$ to 5d $t_{2g}^*$ transition and it is enlarged in the inset, where two peaks labeled IX$_1$ and IX$_2$ are clearly resolved. (b) Several PL spectra measured for different flakes from the sample in Fig. 1 (a). The spectra are collected for different flake sizes ranging from equilateral triangle flakes to a bulk-like flake in addition to very small flakes (SHG).

Fig. 4 (a) A photoluminescence spectrum was measured for a small flake in a sample of ReS$_2$ that was grown at 500 °C with its image shown in the inset. Two peaks observed at E$_1$ = 1.2616 eV and E$_2$ = 2.5418 eV. These two peaks are expanded and shown in the insets. The E$_2$ transition is assigned here as the second harmonic generation of E$_1$ transition. (b) A plot of the electronic band structure of one monolayer of ReS$_2$, left panel, is shown in addition to the projected density of states, right panel, of the Re d-orbital and S p-orbital. The band gap is indicated as 1.43 eV

Fig. 5. An optical absorbance spectrum measured at room temperature for an assembly of flakes grown at silica substrate. An excitonic peak is observed in the absorbance spectrum at around 1.466 eV and compared to an exciton peak measured by the PL at 1.887 eV.

Table I. Raman phone modes in ReS$_2$ measured and calculated with the total modes of 18. The discrepancy between experiment and theory is shown in percentage. Peak number 9 is composed of two peaks originated from bulk and monolayers.

| Peak Number | Experiment (cm$^{-1}$) | Theory (cm$^{-1}$) | Discrepancy (%) |
|---|---|---|---|
| 1 | 131.9 | 129.8 | 1.6 |
| 2 | 138.5 | 137.9 | 0.4 |
| 3 | 148.6 | 148.1 | 0.3 |
| 4 | 159.3 | 158.8 | 0.3 |
| 5 | 210.2 | 210.3 | 0.0 |
| 6 | 233.3 | 230.3 | 1.3 |
| 7 | 273.4 | 262.8 | 3.9 |
| 8 | 280.0 | 269.0 | 3.9 |
| 9 | 304.1 | 296.3 | 2.6 |
| 10 | 316.1 | 304.0 | 3.8 |
| 11 | 322.0 | 311.5 | 3.3 |
| 12 | 344.2 | 333.0 | 3.3 |
| 13 | 368.6 | 354.6 | 3.8 |
| 14 | 372.5 | 363.3 | 2.5 |
| 15 | 405.1 | 393.3 | 2.9 |
| 16 | 416.9 | 406.4 | 2.5 |
| 17 | 432.8 | 424.2 | 2.0 |

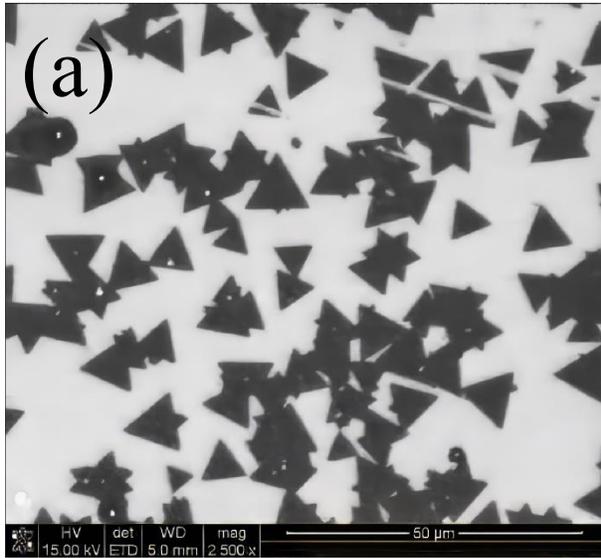
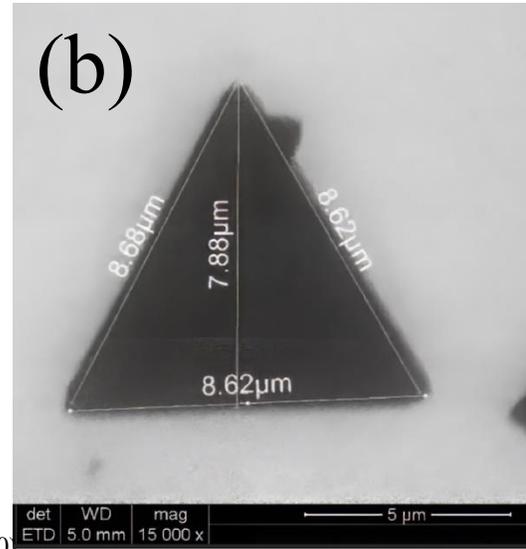
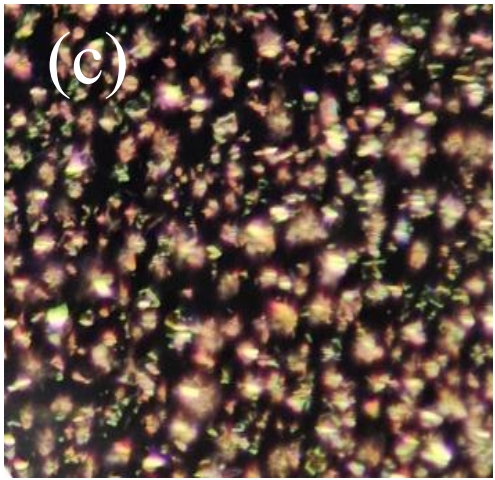
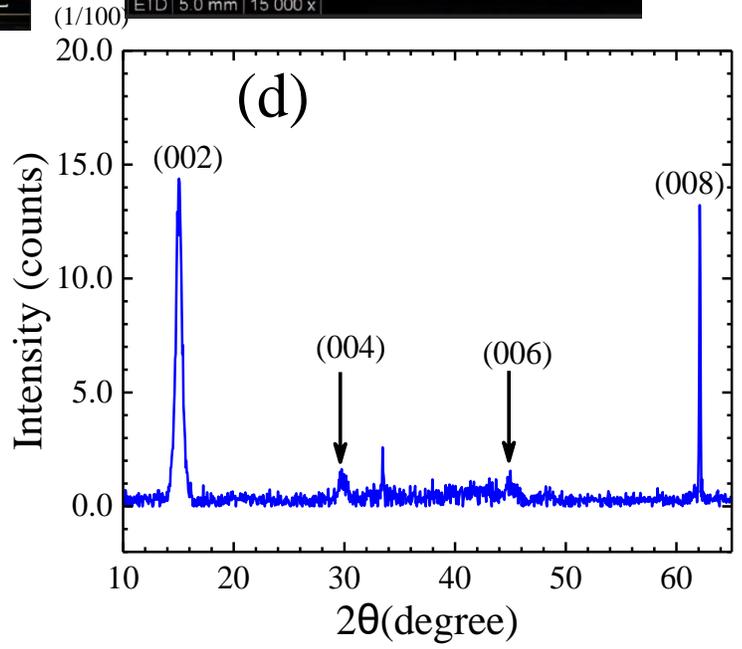

Fig. 1. Ojo, *et al.*

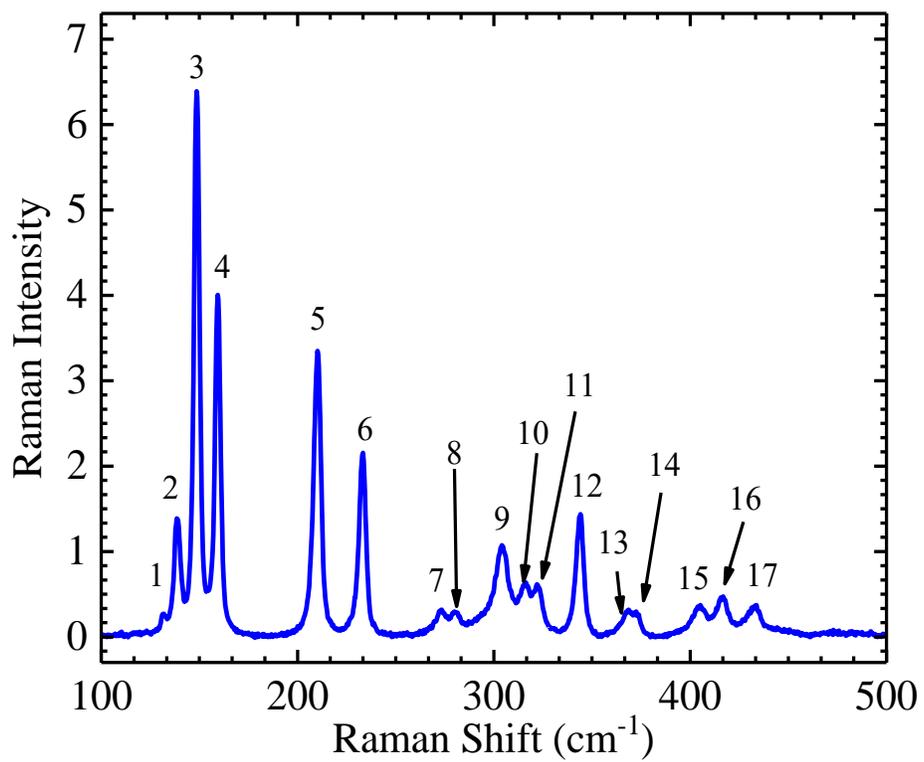

Fig. 2. Ojo, *et al.*

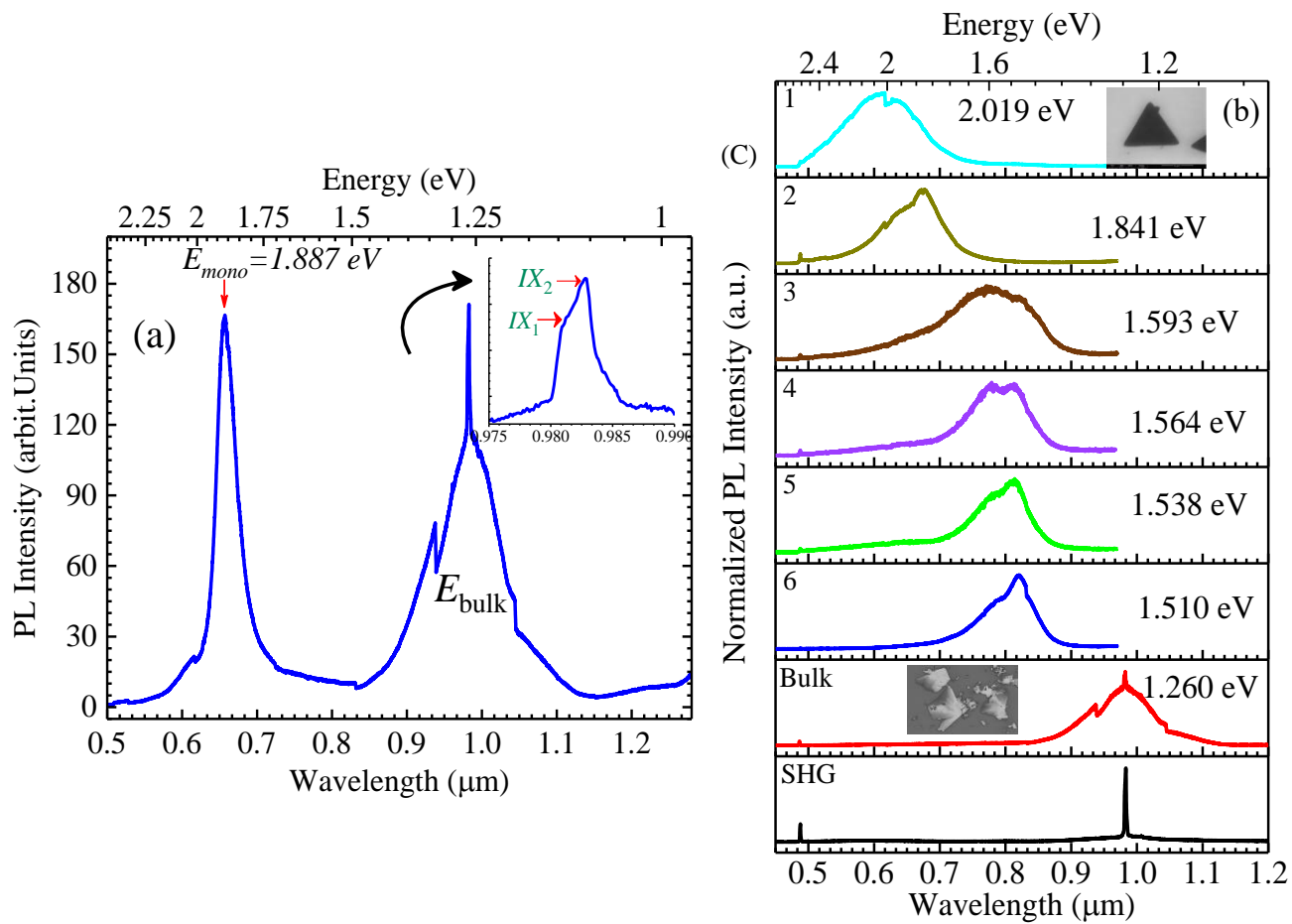

Fig. 3. Ojo, *et al.*

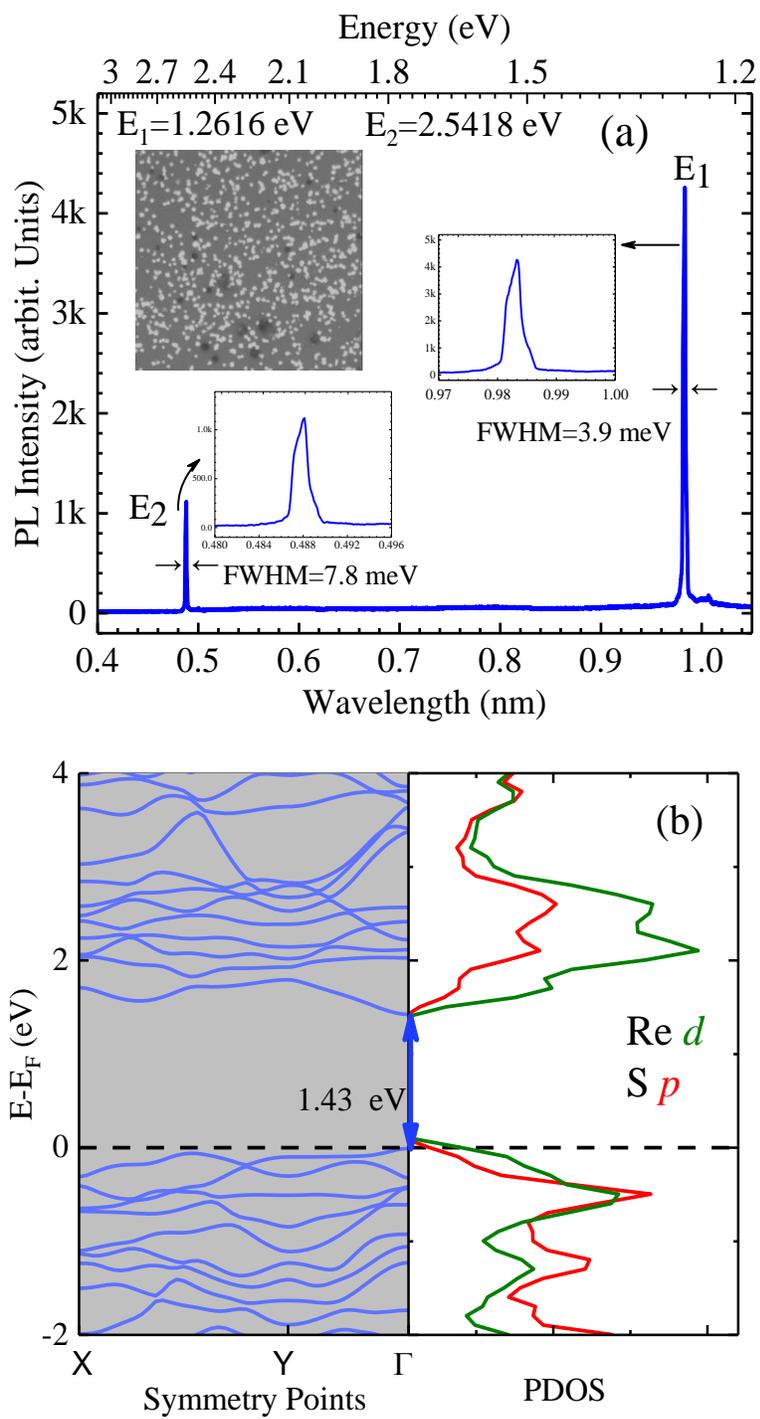

Fig. 4. Ojo, *et al.*

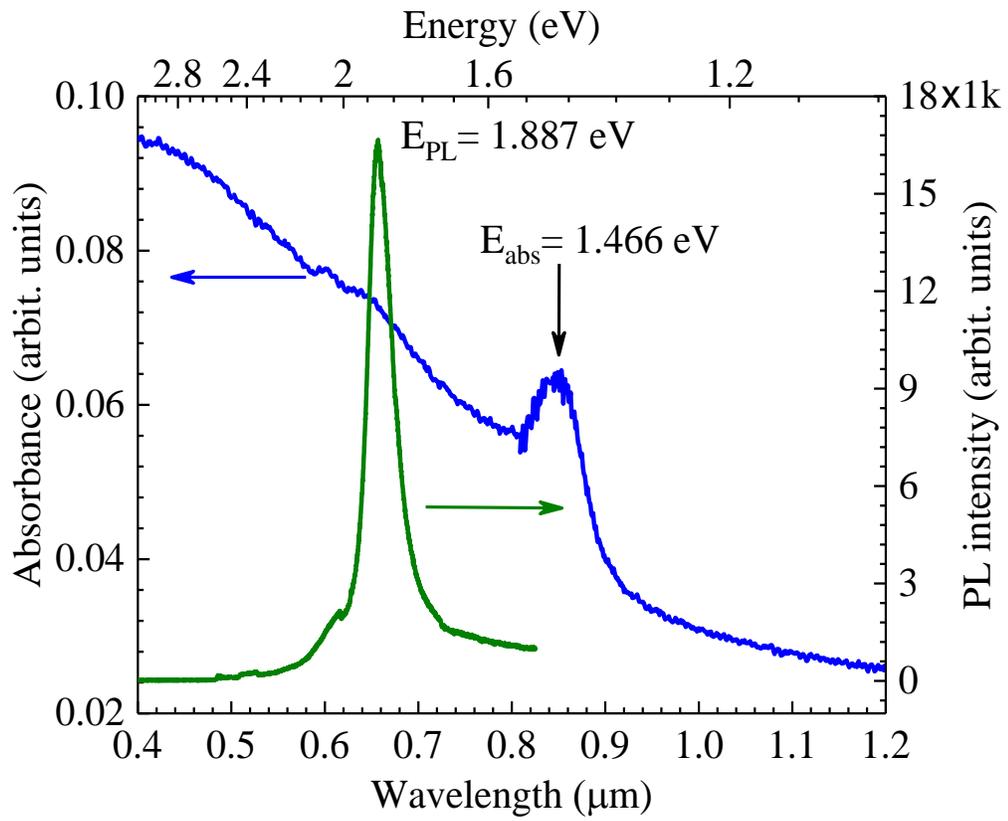

Fig. 5. Ojo, *et al.*